\documentclass[twocolumn,showpacs,preprintnumbers,amsmath,amssymb]{revtex4}

\usepackage{graphicx}
\usepackage{dcolumn}
\usepackage{bm}
\newcommand{\vecvar}[1]{\mbox{\boldmath$#1$}}

\begin{document}

\preprint{PRESAT-8000}

\title{First-principles study on scanning tunneling microscopy images of hydrogen-terminated Si(110) surfaces}

\author{Shinya Horie}
\email{horie@div1.upst.eng.osaka-u.ac.jp}
\affiliation{Applied Science Department, School of Engineering, Osaka University, Suita, Osaka 565-0871, Japan}

\author{Kenta Arima and Kikuji Hirose}
\affiliation{Department of Precision Science and Technology, Osaka University, Suita, Osaka 565-0871, Japan}

\author{Jun Katoh, Tomoya Ono, and Katsuyoshi Endo}
\affiliation{Research Center for Ultra-Precision Science and Technology, Osaka University, Suita, Osaka 565-0871, Japan}

\date{\today}

\begin{abstract}
Scanning tunneling microscopy images of hydrogen-terminated Si(110) surfaces are studied using first-principles calculations. Our results show that the calculated filled-state images and local density of states are consistent with recent experimental results, and the empty-state images appear significantly different from the filled-state ones. To elucidate the origin of this difference, we examined in detail the local density of states, which affects the images, and found that the bonding and antibonding states of surface silicon atoms largely affect the difference between the filled- and empty-state images.
\end{abstract}

\maketitle
As the miniaturization and high integration of electronic devices progress, the atomic and electronic structures of silicon surfaces become areas of intensive research. Scanning tunneling microscopy (STM) \cite{stm} has provided much useful information on surface configurations with an atomic resolution. Many studies revealed that surprisingly complex structures occur in seemingly simple processes such as the adsorption of atoms on semiconductor surfaces. First-principles calculations are also implemented to characterize surface atomic configurations and to interpret electronic structures observed by STM. So far, there have been many studies that reveal interesting atomic and electronic structures on semiconductor surfaces by the combination of experimental and theoretical procedures \cite{ref}.

Nowadays, a Si(110) surface has been considered to be of basic importance in surface science and semiconductor technologies, because the hole mobility of Si(110) with radical oxidation is achieved to be 2.4 times higher than that of Si(001) \cite{so}. However, the atomic and electronic structures of Si(110) surface have not been studied intensively because it is difficult to obtain atomically flat Si(110) surface in experiments. Recently, Arima {\it et al.} observed atomically resolved STM images of hydrogen-terminated Si(110) surfaces after wet cleaning and found certain characteristic features such as a zigzag chain inside a terrace, a single-row bright spot at a step edge and an isolated zigzag chain on a terrace when the sample is rinsed with ultrapure water \cite{ak}.

In this study, we carry out first-principles calculations based on the density-functional theory \cite{hk,ks} to investigate electronic structures of various hydrogen-terminated Si(110) surfaces. Our results show that the calculated STM images of filled states and local density of states (LDOS) are consistent with the results obtained by experiments \cite{ak,ldos}. On the contrary, the empty-state images markedly change from the filled-state ones. The zigzag chains run on the silicon-atom rows of the topmost layers, while they are formed between the silicon-atom rows in the filled-state images. Moreover, the width of the trough composed of a missing monohydride row is greater than that in the filled-state images, and the contribution of the isolated monohydride row to the images is small. These results are explained using the charge distribution of the bonding and antibonding states involved in tunneling.

Our first-principles simulation is based on the real-space finite-difference method \cite{rsfd1,rsfd2,tsdg2,ho1}, which enables us to determine self-consistent electronic ground states and optimized atomic geometries with a high degree of accuracy, using the timesaving double-grid technique \cite{tsdg2,tsdg1,ho1} and the direct minimization of the energy functional \cite{ho}. The norm-conserving pseudopotentials \cite{norm} of Troullier and Martins \cite{tmpp} are employed to describe the electron-ion interaction. Exchange-correlation effects are treated with the local density approximation \cite{lda} in the density-functional theory \cite{hk,ks}. The nine-point finite-difference formula (i.e., the formula with N=4) \cite{rsfd1} is adopted for the derivative arising from the kinetic-energy operator. We take a cutoff energy of 23 Ry, which corresponds to a grid spacing of 0.65 a.u., and a higher cutoff energy of 210 Ry in the vicinity of the nuclei with the augmentation of double-grid points \cite{tsdg2,tsdg1,ho1}.

Figure~\ref{fig:model3} depicts the top views of the computational models adopted here. We employ a technique that involves the use of a supercell whose size is chosen to be $L_x$=61.58 a.u., $L_y$=7.27 a.u., and $L_z$=40.00 a.u., where $L_x$, $L_y$, and $L_z$ are the lengths of the supercell in the $x$, $y$, and $z$ directions, respectively. Here, the direction perpendicular to the surface is chosen to be the $z$ direction. To eliminate completely unfavorable effects of atoms in neighboring cells that are artificially repeated in the case of the periodic boundary condition, the nonperiodic boundary condition of vanishing wave functions out of the supercell is imposed on the $z$ direction, while the periodic boundary condition is adopted in the $x$ and $y$ directions. The supercell contains five silicon layers (i.e., the thin-film model). The topmost layers contain 12, 6, 2, and 10 Si atoms for the (a) terrace, (b) step, (c) isolated monohydride row, and (d) missing monohydride row models in Fig.~\ref{fig:model3}, respectively. The silicon atoms on both surface sides are terminated by hydrogen atoms. Geometry optimization is performed for the three topmost silicon layers until residual forces are less than 0.004 Ry/\AA. The integration over the surface Brillouin zone is implemented using uniform 1$\times$8 $\vecvar{k}$ points. The images on the plane at about 3.1 \AA$\hspace{2mm}$above the topmost silicon-atom layer are generated in the Tersoff-Hamann approximation \cite{th}.

\begin{figure}[bt]
\begin{center}
\vspace{8cm}
\end{center}
\caption{(Color) Top views of topmost and second layers of H-terminated Si(110) surfaces: (a) terrace model, (b) step model, (c) isolated monohydride row model, and (d) missing monohydride row model. Pink and blue balls represent H and Si atoms, respectively, and atoms are denoted by large and small balls according to the distance from the surface.}
\label{fig:model3}
\end{figure}

\begin{figure}[b]
\begin{center}
\vspace{6cm}
\end{center}
\caption{Calculated LDOS on the plane at 3.1 \AA$\hspace{2mm}$above the topmost silicon-atom layer of the atomically flat hydrogen-terminated Si(110) surface. The zero of energy is taken to the Fermi level $E_F$.}
\label{fig:ldos}
\end{figure}
\begin{figure*}[htb]
\begin{center}
\vspace{9cm}
\end{center}
\caption{(Color) Contour plots of charge distributions of LDOS for several energy ranges: (a) $E_F-$1.5 to $E_F-$2.0, (b) $E_F$ to $E_F-$1.5, (c) $E_F$ to $E_F$+1.1, and (d) $E_F$+1.1 to $E_F$+1.5 eV. The contour maps are along the section in the (110) plane (left panel) and the ($\bar{1}$10) plane (right panel). The maps are illustrated on a logarithmic scale, and each contour represents twice or half the density of the adjacent contour lines. Pink and blue balls represent H and Si atoms, respectively, and large (small) balls indicate atomic positions on (below) the cross section.}
\label{fig:ele2}
\end{figure*}

Figures~\ref{fig:ldos} and ~\ref{fig:ele2} show the calculated LDOS of the atomically flat hydrogen-terminated Si(110) surface and the contour plot of charge distributions on the topmost silicon-atom layer for several energy ranges, respectively. The characteristic peaks of the calculated LDOS agree with those obtained by the experiment \cite{ldos}. The filled- and empty-state images are mostly affected by the spatially localized bonding state and nonlocalized antibonding state, respectively, and the peak at 1.25 eV (1.0 eV) below (above) the Fermi level $E_F$ corresponds to the Si-Si and Si-H bonding (antibonding) states. Figure~\ref{fig:STM5} shows the simulated filled- and empty-state images of various hydrogen-terminated Si(110) surfaces. In the filled-state images, the zigzag chains are uniformly oriented along the [$\bar{1}$10] direction, and neighboring hydrogen atoms of two adjacent monohydride rows form a zigzag chain [Fig.~\ref{fig:STM5}(a)]. The hydrogen atoms existing at step edges form single-row chains of bright oval spots [Fig.~\ref{fig:STM5}(b)], and an isolated row on a terrace is composed of a double-row chain of bright protrusions [Fig.~\ref{fig:STM5}(c)]. A missing row consisting of surface silicon and hydrogen atoms is observed to be a dark trough, and its edges are shaped by the single-row chains of bright spots [Fig.~\ref{fig:STM5}(d)]. These images are in agreement with those obtained by the experiment \cite{ak}. On the other hand, the empty-state images appear different from the filled-state images. The zigzag chains exist on the silicon-atom rows of the topmost layer [Figs.~\ref{fig:STM5}(e)-(h)] contrary to the filled-state images in which they are formed between the two adjacent monohydride rows. The monohydrides existing at the step edges appear darker than those on the topmost layer [Fig.~\ref{fig:STM5}(h)]. The width of the dark trough composed of the missing row is observed to be two times greater than that in the filled-state image, and the contribution of the isolated row to the images is small [Fig.~\ref{fig:STM5}(g)].

The difference between the filled- and empty-state images is due to the contributions of the Si-Si bonding and antibonding states involved in tunneling: the filled-state images primarily reflect the spatial distribution of the occupied bonding state, while the empty-state images are affected by the unoccupied antibonding state. In the filled-state images, the positions between the silicon atoms of the topmost layer are dark troughs owing to the small contribution of the spatially localized Si-Si bonding state; therefore, the zigzag chains are observed to be formed by hydrogen atoms of two adjacent monohydride rows. On the other hand, since the Si-Si antibonding state is markedly expanded, the positions between the silicon atoms appear bright and the zigzag chains are formed on the surface silicon-atom rows. The monohydrides existing at the step edges and in the isolated row appear darker than those in the terrace because the Si-H antibonding state of the monohydride at the step edges or in the isolated row shifts higher by $\sim$0.6 eV due to the absence of neighboring monohydride rows.

\begin{figure*}[htb]
\begin{center}
\vspace{13cm}
\end{center}
\caption{(Color) Simulated STM images of H-terminated Si(110) surfaces: filled-state images of (a) terrace, (b) step, (c) isolated row, and (d) missing row models and empty-state images of (e) terrace, (f) step, (g) isolated row, and (h) missing row models. The filled-state images are obtained by integrating charge distributions from $E_F-$2.0 to $E_F$ eV, and the empty-state images, from $E_F$ to $E_F$+1.5 eV. The meanings of the symbols are the same as those in Fig.~\ref{fig:model3}. The equi-LDOS surfaces at a height of about 3.1 \AA $\hspace{2mm}$from the top-most silicon-atom layers are presented.}
\label{fig:STM5}
\end{figure*}

In summary, we have studied the electronic structures of various hydrogen-terminated Si(110) surfaces. The characteristic peaks of the calculated LDOS are consistent with the experimental ones. The formations of zigzag chains and bright protrusions in the filled-state images are in agreement with those obtained by the experiment \cite{ak}. On the other hand, in the empty-state images, zigzag chains shift so as to locate on the silicon-atom rows of the topmost layer. The monohydrides existing at the step edges appear darker than those in the terrace, and those in the isolated row hardly affect the empty-state images. The difference between these images is explained using the contributions of the bonding and antibonding states. To the best of our knowledge, this is the first study that indicates the differences between the filled- and empty-state STM images of the atomically flat hydrogen-terminated Si(110) surface, and an STM study is in progress to verify the results of this investigation.

The authors thank Professor T. Ohmi, Professor S. Sugawa, Associate Professor A. Teramoto, and Dr. H. Akahori of Tohoku University for providing useful information. This research was supported by a Grant-in-Aid for the 21st Century COE ``Center for Atomistic Fabrication Technology'' and also by a Grant-in-Aid for Scientific Research (C) (Grant No. 16605006) from the Ministry of Education, Culture, Sports, Science and Technology. The numerical calculation was carried out by the computer facilities at the Institute for Solid State Physics at the University of Tokyo, and the Information Synergy Center at Tohoku University.


\begin{references}
\bibitem{stm}
G. Binnig, H. Rohrer, Ch. Gerber, and E. Weibel, Phys. Rev. Lett. {\bf 49}, 57 (1982).
\bibitem{ref}
D. Drakova, Rep. Prog. Phys. {\bf 64}, 205 (2001); Werner A. Hofer, Adam S. Foster, and Alexander L. Shluger, Rev. Mod. Phys. {\bf 75}, 1287 (2003), and references therein.
\bibitem{so}
S. Sugawa, I. Ohshima, H. Ishino, Y. Saito, M. Hirayama, and T. Ohmi, IEEE International Electron Devices Meeting Technical Digest, Washington, D.C., December 2-5, 2001, p. 37.3.1-4.
\bibitem{ak}
K. Arima, J. Katoh, and K. Endo, Appl. Phys. Lett. {\bf 85}, 6254 (2004).
\bibitem{hk}
P. Hohenberg and W. Kohn, Phys. Rev. {\bf 136}, B864 (1964).
\bibitem{ks}
W. Kohn and L.J. Sham, Phys. Rev. {\bf 140}, A1133 (1965).
\bibitem{ldos}
M.A. Lutz, R.M. Feenstra, and J.O. Chu, Surf. Sci. {\bf 328}, 215 (1995).
\bibitem{rsfd1}
J.R. Chelikowsky, N. Troullier, and Y. Saad, Phys. Rev. Lett. {\bf 72}, 1240 (1994).
\bibitem{rsfd2}
J.R. Chelikowsky, N. Troullier, K. Wu, and Y. Saad, Phys. Rev. B {\bf 50}, 11355 (1994).
\bibitem{tsdg2}
T. Ono and K. Hirose, cond-mat/0412571 (unpublished).
\bibitem{ho1}
K. Hirose, T.Ono, Y. Fujimoto, and S. Tsukamoto, {\it First-Principles Calculations in Real-Space Formalism, Electronic Configurations and Transport Properties of Nanostructures}, (Imperial College Press, London, 2005).
\bibitem{tsdg1}
T. Ono and K. Hirose, Phys. Rev. Lett. {\bf 82}, 5016 (1999).
\bibitem{ho}
K. Hirose and T. Ono, Phys. Rev. B {\bf 64}, 085105 (2001).
\bibitem{norm}
We used the norm-conserving pseudopotentials NCPS97 constructed by K. Kobayashi. See K. Kobayashi, Comput. Mater. Sci. {\bf 14}, 72 (1999).
\bibitem{tmpp}
N. Troullier and J.L. Martins, Phys. Rev. B {\bf 43}, 1993 (1991).
\bibitem{lda}
J. P. Perdew and A. Zunger, Phys. Rev. B {\bf 23}, 5048 (1981).
\bibitem{th}
J. Tersoff and D.R. Hamann, Phys. Rev. B {\bf 31}, 805 (1985).
\end{references}
\end{document}